\documentclass[a4paper]{article}
\usepackage[latin1]{inputenc} %
\usepackage[T1]{fontenc} %
\usepackage{RR}
\usepackage{hyperref}

\newtheorem{definition}{Definition}

\RRdate{Octobre 2009}
\RRauthor{% les auteurs
N. Malod-Dognin
  \thanks[sfn]{INRIA Rennes - Bretagne Atlantique and University of Rennes 1, France.}%
  \and
R. Andonov%
\thanksref{sfn}
  \and
N. Yanev\thanks{Faculty of Mathematics and Informatics, University of Sofia, Bulgaria.}
	\thanks{Institute of Mathematics and Informatics, Bulgarian Academy of Sciences.}

}
\authorhead{Malod-Dognin, Andonov and Yanev}
\RRtitle{Les cliques maximum dans la comparaison des structures protéiques }
\RRetitle{Maximum Cliques in Protein Structure Comparison}
\titlehead{Maximum Cliques in Protein Structure Comparison}
\RRresume{
Calculer la similarité entre deux structures de protéines est une tâche cruciale de la biologie moléculaire,
et a été étudiée intensément. De nombreuses méthodes de comparaison peuvent être modélisées sous forme de recherches de cliques maximum dans des
graphes $k$-partis spécifiques, que nous appellerons graphes d'alignments.

Dans ce rapport, nous proposons une nouvelle méthode de comparaison de structures protéiques basée sur les distances internes (DAST), qui est formulée
comme une recherche de cliques maximum dans un graphe d'alignement.
Nous avons également concue un algorithme (ACF) pour résoudre de tels problèmes de cliques. ACF est dans un premier temps appliqué dans le contexte
de VAST, un logiciel laregement utilisé au NCBI (National Center for Biotechnology Information), puis il est appliqué dans le contexte de DAST.
Les résultats obtenus sur de véritables instances de comparaison de structures de protéines montrent que notre algorithme est plus de 37000 fois plus rapide
que le solveur original de VAST, qui est basé sur l'algorithme de Bron et Kerbosch. Nous avons ensuite comparé ACF avec l'un des plus rapides algorithmes de recherche de clique maximum, récemment proposé par \"{O}sterg\.{a}rd. Sur un jeu de test connu (l'ensemble de Skolnick), nous observons qu'ACF est en moyenne
20 fois plus rapide que l'algorithme d'\"{O}sterg\.{a}rd.
}

\RRabstract{
Computing the similarity between two protein structures is a crucial task in molecular biology, and has been
extensively investigated. Many protein structure comparison methods can be modeled as maximum clique problems in specific $k$-partite graphs, referred here as alignment  graphs.

In this paper, we propose a new protein structure comparison method based on internal distances (DAST) which is posed as a maximum clique problem in an alignment  graph. We also design an algorithm (ACF) for solving such maximum clique problems. ACF is first applied in the context of VAST, a software largely used in the National Center for Biotechnology Information, and then in the context of DAST. The obtained results on real protein alignment instances show that our algorithm is more than 37000 times faster than the original VAST clique solver which is based on Bron \& Kerbosch algorithm. We furthermore compare ACF with one of the fastest  clique finder, recently conceived by \"{O}sterg\.{a}rd. On a popular benchmark (the Skolnick set) we observe that ACF is about 20 times faster in average than  the \"{O}sterg\.{a}rd's  algorithm. 
}
\RRmotcle{Compraraison de structures protéiques, problème de clique maximum, graphes $k$-partis, optimisation combinatoire.}
\RRkeyword{protein structure comparison, maximum clique problem, $k$-partite graphs, combinatorial optimization.}
\RRprojet{Symbiose}  % cas d'un seul projet
\RRdomaine{1} % cas du domaine numero 1
\RRtheme{Bio}
\URRennes  % pour ceux qui sont \`a l'ouest
\RCRennes % Rennes - Bretagne Atlantique
\begin{document}
\RRNo{7053}
\makeRR   % cas d'un rapport de recherche
%% \makeRT % cas d'un rapport technique.
%% a partir d'ici, chacun fait comme il le souhaite

\section{Introduction}

A fruitful assumption in molecular biology is that proteins of similar three-dimensional
(3D) structures are likely to share a common function and in most cases derive from a same ancestor.
Understanding and computing physical similarity of protein structures is one of the keys for developing protein based
medical treatments, and thus it has been extensively investigated \cite{godzik96,sierk04}.
Evaluating the similarity of two protein structures can be done by finding an optimal (according to some criterions) order-preserving matching (also called alignment) between their components.
We show %in section \ref{alignment} 
that finding such alignments is equivalent to solving maximum clique problems in specific $k$-partite graphs referred here as alignment  graphs. These graphs could be very large (more than 25000 vertices and $3\times 10^7$ edges) when comparing real protein structures.  We are not aware of any previous specialized algorithm for solving  the maximum clique problem in $k$-partite graphs. Even  very recent general clique finders \cite{tomita03,konc07} are  oriented to notably smaller instances and  are not able to solve problems of such  size (the available code of \cite{konc07} is limited to graphs with up to 1000 vertices).

For solving  the maximum clique problem in this context we conceive an algorithm, denoted by \textit{ACF} (for Alignment Clique Finder),  which profits from the particular structure of the alignment  graphs. We furthermore compare ACF to an efficient general clique solver  \cite{ostergard02} and the obtained results clearly demonstrate the usefulness of our dedicated algorithm.

%Let us first introduce the maximum clique problem, and then how protein structure comparison problems can be modeled as
%maximum clique problems in grid graphs.
%
%
\subsection{The maximum clique problem}
We usually denote an undirected graph by $G = (V,E)$, where $V$ is the set of vertices and $E$ is the set of edges.
Two vertices $i$ and $j$ are said to be adjacent if they are connected by an edge of $E$.
A clique of a graph is a subset of its vertex set, such that any two vertices in it are adjacent.
\begin{definition}
The \textbf{maximum clique problem} (also called maximum cardinality clique problem) is to find a largest, in terms of vertices, clique of an
arbitrary undirected graph $G$, which will be denoted by $MCC(G)$.
\end{definition}
The maximum clique problem is one of the first problem shown to be
NP-Complete \cite{karp72} and it has been studied extensively in literature.
Interested readers can refer to \cite{bomze99}
for a detailed state of the art about the maximum clique problem.
%Here we have decided to emphasize on the following applications.
%First, we are not aware of any specialized algorithm for the case
%of $k$-partite graphs. Second, the optimization algorithms proposed so far are not able to solve
%problems of the size necessary in our case. In this study, we compare our algorithm to the \"{O}sterg\.{a}rd algorithm\cite{ostergard02}. Other algorithms have been recently published \cite{tomita03,konc07}, but the available code of \cite{konc07} is limited to relatively small graphs with up to 1000 vertices, while our biggest instances have more than 25000 vertices and $3\times 10^7$ edges.
%Moreover, comparison with ``general'' clique solvers is not fair as they do not take advantage of the partition-structure of our grid graphs.

\subsection{Alignment graphs}
In this paper, we focus on grid alike graphs, which we define as follows.

\begin{definition}\label{def2}
	A $m\times n$ \textbf{alignment graph} $G = (V, E)$ is a graph in which the vertex set $V$ is depicted by a ($m$-rows) $\times$ ($n$-columns) array $T$, where each cell $T[i][k]$ contains at most one vertex $i.k$ from $V$ (note that for both arrays and vertices, the first index stands for the row number, and the second for the column number).
	Two vertices $i.k$ and $j.l$ can be connected by an edge
	$(i.k, j.l) \in E$ only if $i < j$ and $k < l$. An example of such alignment graph is given in Fig \ref{graph_and_order}a.
\end{definition}
It is easily seen that the $m$ rows form a $m$-partition of the alignment graph $G$, and that the $n$ columns also form a $n$-partition.
In the rest of this paper we will use the following notations. A successor of a vertex $i.k \in V$ is an element of the set $\Gamma^+(i.k) = \{ j.l \in V$ s.t. 
$(i.k,j.l) \in E, i<j $ and $k<l \}$. $V^{i.k}$ is the subset of $V$ restricted to vertices in rows $j$, $i \leq j \leq m$, and in columns $l$, $k\leq l \leq n$.
Note that  $\Gamma^+(i.k) \subset V^{i+1.k+1}$. $G^{i.k}$ is the subgraph of $G$ induced by the vertices in $V^{i.k}$.
The cardinality of
a vertex set $U$ is $|U|$.

\subsection{Relations with protein structure similarity}\label{alignment}
From a general point of view, two proteins $P_1$ and $P_2$ can be represented by their ordered set of components $N_1$ and $N_2$, and estimating their similarity can be done by finding the
longest
alignment between the elements of $N_1$ and $N_2$.
In our approach, such matchings are represented in a $|N_1|\times |N_2|$ alignment graph $G=(V,E)$, where each row corresponds
to an element of $N_1$ and each column corresponds to an element of $N_2$. A vertex $i.k$ is in $V$ (i.e. matching $i\leftrightarrow k$ is possible),
only if element $i\in N_1$ and $k \in N_2$ are compatible.
An edge $(i.k,j.l)$ is in  $E$ if and only if (i) $i<j$ and $k<l$, for order preserving, and (ii) matching $i\leftrightarrow k$ is compatible with matching $j\leftrightarrow l$.
A feasible matching of $P_1$ and $P_2$ is then a clique in $G$, and the longest alignment corresponds to a maximum clique in $G$.
%Between many alignment methods, differences lay both in the nature of the elements of $N_1$ and $N_2$ and in the compatibility definitions between elements and between pairs of matched elements.
There is a multitude of alignment methods and they differ mainly by the nature of the elements of $N_1$ and $N_2$ and by the compatibility definitions between elements and between pairs of matched elements.
At least two protein structure similarity related problems from the literature can be converted into clique problems in alignment graphs :
the secondary structure alignment in VAST\cite{gibrat96}, and the Contact Map Overlap Maximization problem (CMO)\cite{godzik94}.

\textbf{VAST}, or Vector Alignment Search Tool, is a software for aligning protein 3D structures largely used in the National Center for Biotechnology Information
\footnote{http://www.ncbi.nlm.nih.gov/Structure/VAST/vast.shtml}.
In VAST, $N_1$ and $N_2$  contain 3D vectors representing the secondary structure elements (SSE) of $P_1$ and $P_2$.
Matching $i\leftrightarrow k$ is possible if vectors $i$ and $k$ have similar norms and correspond either both to $\alpha$-helices or both to $\beta$-strands.
Finally, matching $i\leftrightarrow k$ is compatible with matching $j\leftrightarrow l$ only if the couple of vectors $(i, j)$ from $P_1$
can be well superimposed in 3D-space with the couple of vectors $(k, l)$ from $P_2$.

\textbf{CMO} is one of the most reliable and robust measures of protein structure similarity. Comparisons are done by aligning the residues (amino-acids) of two proteins in a way that maximizes the number of common contacts (when two residues that are close in 3D space are matched with two residues that are also close in 3D space). We have already dealt with CMO in \cite{wabi08}, but not by using cliques. Note that a maximum clique formulation in alignment graphs was proposed by Strickland et al. in \cite{strickland05}, but this formulation differs from ours.

\subsection{DAST: an improvement of CMO based on internal distances}
\begin{figure}[!h]
	\footnotesize
	\caption{An optimal CMO matching.}\label{cmo_matching}
	\begin{center}
		\includegraphics[width=2.5cm]{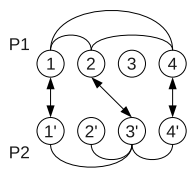}
	\end{center}
	Two proteins ( $P_1$ and $P_2$) are represented by their contact map  graphs where the vertices corresponds to the residues and where edges connect residues in contacts (i.e. close).
	The matching ``$1\leftrightarrow 1', 2\leftrightarrow 3', 4\leftrightarrow 4'$'', represented by the arrows, yields two common contacts which is the maximum for the considered case. However, it also matches residues $1$ and $4$ from $P_1$ which are in contacts with residues $1'$ and $4'$ in $P_2$ which are remote.
\end{figure}

One of the main drawback of CMO is that in order to maximize the number of common contacts, it also introduces some ``errors'' like aligning two residues that are close in 3D space with two residues that are remote, as illustrated in Fig \ref{cmo_matching}. These  errors could potentially yield alignments with big root mean square deviations (RMSD) which is not desirable for structures comparison.
To avoid such problems we propose DAST (Distance-based Alignment Search Tool), an alignment method based on internal distances which is modeled in an alignment graph. In DAST, the two proteins $P_1$ and $P_2$ are represented by their ordered sets of residues $N_1$ and $N_2$.
Two residues $i\in N_1$ and $k\in N_2$ are compatible if they come from the same kind of secondary structure elements (i.e. $i$ and $k$ both come from an $\alpha$-helix, or from a $\beta$-strand) or if both come from a loop.
Let us denote by $d_{ij}$ (resp. $d_{k.l}$) the euclidean distance between the $\alpha$-carbons of residues $i$ and $j$ (resp. $k$ and $l$).
Matching $i\leftrightarrow k$ is compatible with matching $j\leftrightarrow l$ only if $|d_{ij} - d_{kl}| \leq \tau$, where $\tau$ is a distance threshold.
The longest alignment in terms of residues, in which each couple of residues from $P_1$ is matched with a couple of residues from $P_2$ having similar distance relations, corresponds to a maximum clique in $G$.
Since $RMSD = \sqrt{ \frac{1}{N_m} \times \sum(|d_{ij}-d_{kl}|^2)} $, where $N_m$ is the number of matching pairs ``$i\leftrightarrow k, j\leftrightarrow l$'',
the alignments given by DAST have a RMSD of internal distances $\leq \tau$.

\newpage

\section{Branch and Bound approach}
We have been inspired by \cite{ostergard02} to propose our own algorithm which is more suitable for solving the maximum clique problem in the previously
defined $m\times n$ alignment graph $G=(V,E)$.
Let $Best$ be the biggest clique found so far (first it is set to $\emptyset$), and ${|\overline{MCC}(G)|}$ be an over-estimation of $|MCC(G)|$.
By definition, $V^{i+1.k+1} \subset V^{i.k+1} \subset V^{i.k}$, and similarly $V^{i+1.k+1} \subset V^{i+1.k} \subset V^{i.k}$.
From these inclusions and from definition\ref{def2}, it is easily seen that for any $G^{i.k}$, $MCC(G^{i.k})$ is the biggest clique among $MCC(G^{i+1.k})$, $MCC(G^{i.k+1})$ and $MCC(G^{i+1.k+1})$ $\bigcup$ $\{i.k\}$, but for the latter only if vertex $i.k$ is adjacent to all vertices in $MCC(G^{i+1.k+1})$.
Let $C$ be a $(m+1) \times (n+1)$ array where $C[i][k] = |\overline{MCC}(G^{i.k})|$ (values in row $m+1$ or column $n+1$ are equal to 0).
For reasoning purpose, let assume that the upper-bounds in $C$ are exact.
If a vertex $i.k$ is adjacent to all vertices in $MCC(G^{i+1.k+1})$, then $C[i][k]$ = $1 + C[i+1][k+1]$, else $C[i][k] = \max (C[i][k+1]$, $C[i+1][k])$.
We can deduce that a vertex $i.k$ cannot be in a clique in $G^{i.k}$ which is bigger than $Best$ if $C[i+1][k+1] < |Best|$, and this reasoning still holds if values in $C$ are upper estimations.
Another important inclusion is $\Gamma^+(i.k) \subset V^{i+1.k+1}$. Even if $C[i+1][k+1] \geq |Best|$, if $|\overline{MCC}(\Gamma^+(i.k))| < |Best|$ then $i.k$ cannot be in a clique in $G^{i.k}$ bigger than $Best$.

\begin{figure}[!h]
	\footnotesize
	\caption{A $4\times 4$ alignment graph and the visiting order of its array $T$}\label{graph_and_order}
	\begin{center}
		\includegraphics[width=12cm]{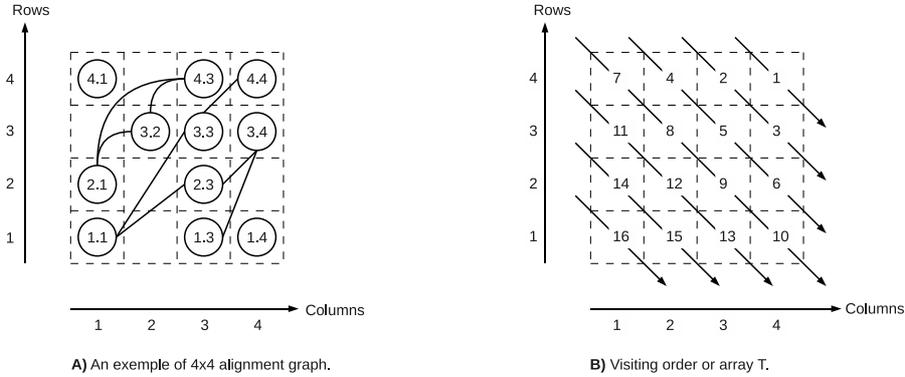}
	\end{center}
\end{figure}

Our main clique cardinality estimator is constructed and used according to these properties. A function, Find\_clique($G$), will visit the cells of $T$ according to north-west to south-est diagonals, from diagonal ``$i+k = m+n$'' to diagonal ``$i+k = 2$'' as illustrated in Fig \ref{graph_and_order}b. For each cell $T[i][k]$ containing a vertex $i.k\in V$, it may call Extend\_clique($\{i.k\}$, $\Gamma^+(i.k)$), a function which tries to extend the clique $\{i.k\}$ with vertices in $\Gamma^+(i.k)$ in order to obtain a clique bigger than $Best$ (which cannot be bigger than |Best| +1). If such a clique is found, $Best$ is updated.
However, Find\_clique() will call Extend\_clique() only if two conditions are satisfied : (i) $C[i+1][k+1] = |Best|$ and (ii) $|\overline{MCC}(\Gamma^+(i.k))| \geq |Best|$. After the call to Extend\_clique(), $C[i][k]$ is set to $|Best|$.
For all other cells $T[i][k]$, $C[i][k]$ is set to $\max (C[i][k+1]$, $C[i+1][k])$ if $i.k \notin V$, or to $1 + C[i+1][k+1])$ if $i.k \in V$.
Note that the order used for visiting the cells in $T$ guaranties that when computing the value of $C[i][k]$, the values of $C[i+1][k]$, $C[i][k+1]$ and $C[i+1][k+1]$ are already computed.

Array $C$ can also be used in function Extend\_clique() to fasten the maximum clique search.
This function is a branch a bound (B\&B) search using the following branching rules.
Each node of the B\&B tree is characterized by a couple ($Cli$, $Cand$) where $Cli$ is the clique under construction and $Cand$ is the set of candidate vertices to be added to $Cli$.
Each call to Extend\_clique($\{i.k\}$, $\Gamma^+(i.k)$) create a new B\&B tree which root node is ($\{i.k\}$, $\Gamma^+(i.k)$).
The successors of a B\&B node $(Cli, Cand)$ are the nodes $(Cli \bigcup \{i'.k'\}$, $Cand \bigcap \Gamma^+(i'.k'))$, for all vertices $i'.k' \in Cand$. Branching follows lexicographic increasing order (row first).
According to the branching rules, for any given B\&B node ($Cli$, $Cand$) the following cutting rules holds :
(i) if $|Cli|$ + $|Cand| \leq |Best|$ then the current branch cannot lead to a clique bigger than $|Best|$ and can be fathomed, 
(ii) if $|\overline{MCC}(Cand)| \leq |Best| - |Cli|$, then the current branch cannot lead to a clique bigger than $|Best|$,
and (iii) if $|\overline{MCC}(Cand \bigcap \Gamma^+(i.k))| \leq |Best| - |Cli| -1$, then branching on $i.k$ cannot lead to a clique bigger than $|Best|$.
For any set $Cand$ and any vertex $i.k$, $Cand \bigcap \Gamma^+(i.k) \subset \Gamma^+(i.k)$ , and $\Gamma^+(i.k) \subset G^{i+1.k+1}$. From these inclusions we can deduce two way of over-estimating $|MCC(Cand \bigcap \Gamma^+(i.k))|$. First, by using $C[i+1][k+1]$ which over-estimate $|MCC(G^{i+1.k+1})|$
and second, by over-estimating $|MCC(\Gamma^+(i.k))|$. All values $|\overline{MCC}(\Gamma^+(i.k))|$ are computed once for all in Find\_clique() and thus,
only $|\overline{MCC}(Cand)|$ needs to be computed in each B\&B node.

\newpage

\section{Maximum clique cardinality estimators}\label{estimators}

Even if the described functions depend on array $C$, they also use another upper-estimator of the cardinality of a maximum clique in an alignment graph.
By using the properties of alignment graphs, we developed the following estimators.

\subsection{Minimum number of rows and columns}
Definition \ref{def2} implies that there is no edge between vertices from the same row or the same column. This means that in a $m\times n$ alignment graph,
$|MCC(G)| \leq \min(m,n)$.
If the numbers of rows and columns are not computed at the creation of the alignment graph, they can be computed in $O(|V|)$.

\subsection{Longest increasing subset of vertices}

\begin{definition}
An \textbf{increasing subset of vertices} in an alignment graph $G=\{V, E\}$ is an ordered subset $\{ i_1.k_1$, $i_2.k_2$, $\ldots$, $i_t.k_t$ \} of $V$, such that $\forall j\in [1,t-1]$, $i_j < i_{j+1}$, $k_j < k_{j+1}$. $LIS(G)$ is the longest, in terms of vertices, increasing subset of vertices of $G$.
\end{definition}

Since any two vertices in a clique are adjacent, definition \ref{def2} implies that a clique in $G$ is an increasing subset of vertices. However, an increasing subset of vertices is not necessarily a clique (since vertices are not necessarily adjacent), and thus $|MCC(G)| \leq |LIS(G)|$.
In a $m \times n$ alignment graph $G = (V,E)$, $LIS(G)$ can be computed in $O(n \times m)$ times by dynamic programming.
However, it is possible by using the longest increasing subsequence to solve $LIS(G)$ in $O(|V|\times \ln(|V|) )$ times which is more suited
in the case of sparse graph like in our protein structure comparison experiments.

\begin{definition}
The \textbf{longest increasing subsequence} of an arbitrary finite sequence of integers $S =$ ``$i_i, i_2, \ldots, i_n$'' is 
the longest subsequence $S'=$ ``$i'_i, i'_2, \ldots, i'_t$'' of $S$ respecting the original order of $S$, and such that for all $j\in [1, t], i'_j < i'_{j+1}$.
By example, the longest increasing subsequence of ``1,5,2,3'' is ``1,2,3''.
\end{definition}

For any given alignment graph $G=\{V, E\}$, we can easily reorder the vertex set $V$, first by increasing order of columns, and second by decreasing order of rows.
Let's denote by $V'$ this reordered vertex set. Then we can create an integer sequence $S$ corresponding to the row indexes of vertices in $V'$.
For example, by using the alignment graph presented in Fig\ref{graph_and_order}a, the reordered vertex set $V'$ is $\{4.1$, $~2.1$, $~1.1$, $~3.2$,$~ 4.3$,$ ~ 3.3$, $~ 2.3$, $~1.3$, $~4.4$, $~3.4$, $~1.4\}$, and the corresponding sequence of row indexes $S$ is ``$4$, $~2$, $~1$, $~3$, $~4$, $~3$, $~2$, $~1$, $~4$, $~3$, $~1$''.
An increasing subsequence of $S$ will pick at most one number from a column, and thus an increasing subsequence is longest if and only if it covers a maximal number
of increasing rows.
This proves that solving the longest increasing subsequence in $S$ is equivalent to solving the longest increasing subset of vertices in $G$.
Note that the longest increasing subsequence problem is solvable in time $O(l\times \ln(l) )$ \cite{fredman75}, where $l$ denotes the length of the input sequence. In our case, this corresponds to $O(|V|\times \ln(|V|) )$.

\subsection{Longest increasing path}
\begin{definition}
An \textbf{increasing path} in an alignment $G=\{V, E\}$ is an increasing subset of vertex \{$i_1.k_1$, $i_2.k_2$, $\ldots$, $i_t.k_t\}$ such that $\forall j\in [1,t-1]$, $(i_j.k_j,i_{j+1}.k_{j+1}) \in E$. The longest increasing path in $G$ is denoted by $LIP(G)$
\end{definition}
As the increasing path take into account edges between consecutive vertices, $|LIP(G)|$, should better estimate $MCC(G)|$.
$|LIP(G)|$ can be computed in $O(|V|^2)$ by the following recurrence. Let $DP[i][k]$ be the length of the longest increasing path in $G^{i.k}$ containing vertex $i.k$.
$DP[i][k] = 1 +$ $\max_{i'.k' \in \Gamma^+{i.k}} ( DP[i'][k'] )$.
The sum over all $\Gamma^+(i.k))$ is done in $O(|E|)$ time complexity, and finding the maximum over all $DP[i][k]$ is done in $O(|V|)$.
This results in a $O(|V| + |E|)$ time complexity for computing $|LIP(G)|$.

Amongst all of the previously defined estimators, the longest increasing subset of vertices (solved using the longest increasing subsequence) exhibits the best performances and is the
one we used for obtaining the results presented in the next section.

\newpage

\section{Results}
All results presented in this section come from real protein structure comparison instances.
Our algorithm, denoted by \textit{ACF} (for Alignment Clique Finder), has been implemented in C and was tested in two different contexts:  secondary structure alignments in VAST and  residue alignments in DAST.
\textit{ACF} will be compared to \"{O}sterg\.{a}rd's algorithm\cite{ostergard02} (denoted by \textit{\"{O}sterg\.{a}rd}) and to the original VAST clique solver which is based on Bron and Kerbosch's algorithm\cite{bron73} (denoted by \textit{BK}). Note that \textit{BK} is not a maximum clique finder but returns all maximal cliques in a graph.

\subsection{Secondary structures alignments}\label{sec1}

This section illustrates the  behavior of \textit{ACF}  in the context of secondary structure element (SSE) alignments. For this purpose we integrated \textit{ACF} and  \textit{\"{O}sterg\.{a}rd} (which code is freely available) in VAST. We afterwards compared them with \textit{BK} by selecting few large protein chains having between 80 to 90 SSE's  (for smaller protein chains the running times of both \textit{\"{O}sterg\.{a}rd} and \textit{ACF} are less than 0.01 sec.).
Computations were done on a AMD at 2.4 GHz computer, and the corresponding running times are presented in \textbf{table \ref{sse_results}}. We observe that 
\textit{\"{O}sterg\.{a}rd} is 4053 times faster than \textit{BK}, and that \textit{ACF} is about 9.3 times faster than \textit{\"{O}sterg\.{a}rd}. Although we have chosen large protein chains, the SSE alignment graphs are relatively small (up to 5423 vertices and 551792 edges ).  On such graphs the difference  between \textit{\"{O}sterg\.{a}rd} and \textit{ACF} performance  is not very  visible--it will be better illustrated on larger alignment  graphs in the next  section.

\begin{table}[!h]
\footnotesize
	\caption{Runing time comparison on secondary structure  alignment  instances}
	\label{sse_results}
	\begin{center}
		\begin{tabular}{|l l| c c c|}
			\hline
			\multicolumn{2}{|c|}{Instances}	& ~BK (sec.)~		& ~\"Osterg\.{a}rd (sec.)~	& ~ACF (sec.)~ \\
			\hline
			~1k32B	& 1n6eI~			& 1591.89		& 1.42				& \bf{0.09} \\
			~1k32B	& 1n6fB~			& 1546.78		& \bf{0.01}			& \bf{0.01} \\
			~1k32B	& 1n6fF~			& 1584.25		& 0.14				& \bf{0.02} \\
			~1n6dD	& 1k32B~			& 1373.35		& 0.06				& \bf{0.01} \\
			~1n6dD	& 1n6eI~			& 1390.27		& 0.11				& \bf{0.03} \\
			~1n6dD	& 1n6fB~			& 1328.85		& 0.65				& \bf{0.06} \\
			~1n6dD	& 1n6fF~			& 1398.41		& 0.13				& \bf{0.05} \\
			\hline
		\end{tabular}
	\end{center}
	Runing time comparison of \textit{BK}, \textit{\"{O}sterg\.{a}rd} and \textit{ACF} on secondary structure alignment instances
	for long protein chains (containing from 80 to 90 SSE's). \textit{BK} is notably slower than the \textit{\"{O}sterg\.{a}rd}'s  algorithm, which is  slightly  slower than \textit{ACF}.
\end{table}

\subsection{Residues alignment}\label{res_align}\label{sec2}
In this section we compare \textit{ACF} to \textit{\"{O}sterg\.{a}rd} in the context of residue alignments in DAST.
Computations were done on a PC with an Intel Core2 processor at 3Ghz, and for both algorithms the computation time was bounded to 5 hours per instance. Secondary structures assignments were done by KAKSI\cite{kaksi}, and the  threshold distance $\tau$ was set to 3\AA. The protein structures come from the well known Skolnick set, described in \cite{lancia01}. It contains 40 protein chains having from 90 to 256 residues, classified in SCOP\cite{murzin} (v1.73) into five families.
Amongst the 780 corresponding alignment instances, 164 align protein chains from the same family and will be called ``similar''. The 616 other instances align protein chains from different families and thus will be called ``dissimilar''.
Characteristics of the corresponding alignment graphs are presented in \textbf{table \ref{residues_graphs}}.

\begin{table}[!h]
\footnotesize
	\caption{DAST alignment graphs characteristics}
	\begin{center}
		\label{residues_graphs}
		\begin{tabular}{|l|c|c|c|c|c|c|}
			\cline{3-7}
			\multicolumn{2}{c|}{~}	& ~array size~		& ~|V|~		&  ~|E|~	& ~density~	& ~|MCC|~\\
			\hline
			~similar~	&~min~	& ~97$\times$97~	& ~4018~	& ~106373~	& ~8.32\%~	& ~45~\\
			\cline{2-7}
			~instances~	&~max~	& 256$\times$255~	& ~25706~	& ~31726150~	& ~15.44\%~	& ~233~\\
			\hline
			~dissimilar~	&~min~	& ~97$\times$104~	& ~1581~	& ~77164~	& ~5.76\%~	& ~12~\\
			\cline{2-7}
			~instances~	&~max	& ~256$\times$191~	& ~21244~	& ~16839653~	& ~14.13\%~	& ~48~\\
			\hline
			\end{tabular}
	\end{center}
	All alignment  graphs from DAST have small edge density (less than 16\%). Similar instances are characterized by bigger maximum cliques than the
	dissimilar instances.
\end{table}

\textbf{Table \ref{Solved_instances}} compares the number of instances solved  by each algorithm on 
Skolnick set. \textit{ACF} solved 155 from 164 similar instances, while \textit{\"{O}sterg\.{a}rd}  solved 128 instances.
\textit{ACF} was able to solve all 616 dissimilar instances, while \textit{\"{O}sterg\.{a}rd}  solved 545 instances only.
Thus, on this popular benchmark set,  \textit{ACF} clearly outperformed \textit{\"{O}sterg\.{a}rd} in terms of number of solved instances.

\begin{table}[!h]
\footnotesize
	\begin{center}
		\caption{Number of solved instances comparison }
		\label{Solved_instances}
		\begin{tabular}{|r|c c|}
			\cline{2-3}
			\multicolumn{1}{c|}{~}		&~\"{O}sterg\.{a}rd~	& ~ACF~\\
			\hline
			~Similar instances (164)~	& 128			& ~{\bf 155}~\\
			~Dissimilar instances (616)~	& 545			& ~{\bf 616}~\\
			~Total (780)~           	& 673			& ~{\bf 771}~\\
			\hline
		\end{tabular}
	\end{center}
	Number of solved instances  on Skolnick set: \textit{ACF} solves 21\% more similar instances and 13\% more dissimilar instances than \textit{\"{O}sterg\.{a}rd}.
\end{table}

\begin{figure}[!h]
\footnotesize
	\begin{center}
		\caption{Running time comparison on the Skolnick set}\label{res_time_all}
		\includegraphics[angle=270,width=7cm]{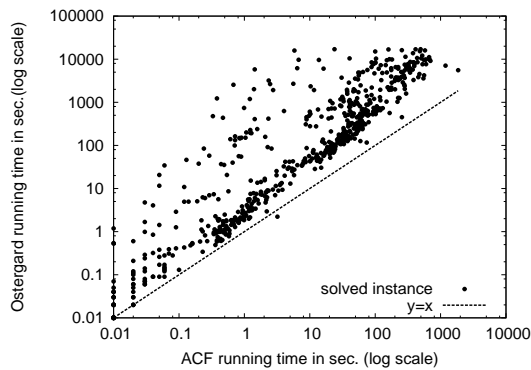}
	\end{center}
	\textit{ACF} versus \textit{\"{O}sterg\.{a}rd} running time comparison on the set of the 673 Skolnick instances
	solved by both algorithms. The \textit{ACF} time is presented on the x-axis, while the
	one of \textit{\"{O}sterg\.{a}rd} is on the y-axis. For all instances except one, \textit{ACF} is faster than \textit{\"{O}sterg\.{a}rd}.
\end{figure}

\textbf{Figure~\ref{res_time_all}} compares the running time of \textit{ACF} to the one of \textit{\"{O}sterg\.{a}rd} on the set of 673 instances solved by both algorithms (all instances solved by \textit{\"{O}sterg\.{a}rd} were also solved by \textit{ACF}).
For all instances except one, \textit{ACF} is significantly faster than \textit{\"{O}sterg\.{a}rd}.
More precisely, \textit{ACF} needed 12 hs. 29 min. 56 sec. to solve all these 673 instances, while \textit{\"{O}sterg\.{a}rd} needed 260 hs. 10 min. 10 sec. Thus, on the Skolnick set, \textit{ACF} is about 20 times faster in average  than \textit{\"{O}sterg\.{a}rd}, (up to 4029 times for some intstances).

\newpage
\section{Conclusion and future work}

In this paper we introduce a novel protein structure comparison approach  DAST, for Distance-based Alignment Search Tool.
% It is based on the distance between the residues, and for any fixed threshold $\tau$,
% aligns the maximum number of pairs, such that the RMSD of all aligned pairs is no more than $t$. 
For any fixed threshold $\tau$, it finds the longest alignment in which each couple of pairs of matched residues shares the same distance relation (+/- $\tau$), and thus the RMSD of the alignment is $\leq \tau$.
This property is not guaranteed  by the CMO approach, which inspired initially DAST.
From computation standpoint, DAST requires solving  the  maximum clique problem in a specific $k$-partite graph. By exploiting the peculiar structure of this graph, we design  a new maximum clique solver which significantly outperforms 
one of the best general  maximum clique solver.   Our solver was successfully integrated into two protein structure comparison softwares and will be freely available  soon.  We are currently studying  the  quality of DAST alignments from practical viewpoint and  compare the obtained  results with other structure comparison methods.  

\newpage
\section*{Acknowledgements}
This work is a part of ANR project PROTEUS ``ANR-06-CIS6-008'',
No\"el Malod-Dognin is supported by the Brittany Region, and
Nicola Yanev is supported by the bulgarian project DVU/01/197, South-West University, Blagoevgrad.
All computations were done on the Ouest-genopole bioinformatics platform (http://genouest.org).
We would like to express our gratitude to J-F Gibrat for numerous helpful discussions and for providing us
the source code of VAST.

\newpage

\bibliographystyle{plain}
\bibliography{biblio}

\newpage

\tableofcontents

\end{document}